\documentclass[english,amsmath,amssymb,superscriptaddress,prl,nofootinbib]{revtex4-1}
\usepackage[T1]{fontenc}
\usepackage[latin9]{inputenc}
\usepackage{geometry}
\usepackage{color}
\definecolor{note_fontcolor}{rgb}{0.80078125, 0.80078125, 0.80078125}
\usepackage{units}
\usepackage{textcomp}
\usepackage{amsmath}
\usepackage{amssymb}
\usepackage{graphicx}
\usepackage{esint}
\PassOptionsToPackage{normalem}{ulem}
\usepackage{ulem}

\makeatletter

\providecommand{\tabularnewline}{\\}
\newenvironment{lyxgreyedout}
  {\textcolor{note_fontcolor}\bgroup\ignorespaces}
  {\ignorespacesafterend\egroup}

\@ifundefined{textcolor}{}
{%
 \definecolor{BLACK}{gray}{0}
 \definecolor{WHITE}{gray}{1}
 \definecolor{RED}{rgb}{1,0,0}
 \definecolor{GREEN}{rgb}{0,1,0}
 \definecolor{BLUE}{rgb}{0,0,1}
 \definecolor{CYAN}{cmyk}{1,0,0,0}
 \definecolor{MAGENTA}{cmyk}{0,1,0,0}
 \definecolor{YELLOW}{cmyk}{0,0,1,0}
 }

\usepackage{amsthm}
\newcommand{\asterism}{\smash{%
\raisebox{-.5ex}{%
\setlength{\tabcolsep}{-.5pt}%
\begin{tabular}{@{}cc@{}}%
\multicolumn2c*\\[-2ex]*&*%
\end{tabular}}}}

\makeatother

\usepackage{babel}
\begin{document}

\title{Analytic influence functionals for numerical Feynman integrals in
most open quantum systems}

\author{Nikesh S. Dattani}

\email{nike.dattani@chem.ox.ac.uk}

\affiliation{Physical and Theoretical Chemistry Laboratory, Department of Chemistry,
University of Oxford, Oxford, OX1 3QZ, UK}

\author{Felix A. Pollock}

\email{felix.pollock@physics.ox.ac.uk}

\affiliation{Clarendon Laboratory, Department of Physics, University of Oxford,
Oxford, OX1 3PU, UK}

\author{David M. Wilkins}

\email{david.wilkins@seh.ox.ac.uk}

\affiliation{Physical and Theoretical Chemistry Laboratory, Department of Chemistry,
University of Oxford, Oxford, OX1 3QZ, UK}

\date{14 March 2012}
\begin{abstract}
Fully analytic formulas, which do not involve any numerical integration,
are derived for the discretized influence functionals of a very extensive
assortment of spectral distributions. For Feynman integrals derived
using the Trotter splitting and Strang splitting, we present general
formulas for the discretized influence functionals in terms of proper
integrals of the bath response function. When an analytic expression
exists for the bath response function, these integrals can almost
always be evaluated analytically. In cases where these proper integrals
cannot be integrated analytically, numerically computing them is much
faster and less error-prone than calculating the discretized influence
functionals in the traditional way, which involves numerically calculating
integrals whose bounds are both infinite. As an example, we present
the analytic discretized influence functional for a bath response
function of the form $\alpha(t)=\sum_{j}^{K}p_{j}e^{\Omega_{j}t}$,
which is a natural form for many spectral distribution functions (including
the very popular Lorentz-Drude/Debye function), and for other spectral
distribution functions it is a form that is easily obtainable by a
least-squares fit. Evaluating our analytic formulas for this example
case is much faster and more rigorous than numerically calculating
the discretized influence functional in the traditional way. In the
appendix we provide analytic expressions for $p_{j}$ and $\Omega_{j}$
for a variety of spectral distribution forms, and as a second example
we provide the analytic bath response function and analytic influence
functionals for spectral distributions of the form $J(\omega)\propto\omega^{s}e^{-(\omega/\omega_{c})^{q}}.$
The value of the analytic expression for this bath response function
extends beyond its use for calculating Feynman integrals. We also
provide open source MATLAB and Mathematica programs to make the results
of this paper very easy to implement.
\end{abstract}
\maketitle

\section{Introduction}

Influence functionals are used to describe the environment's influence
on the reduced density operator dynamics (\textbf{R$\rho$D}) of open
quantum systems (\textbf{OQS}s) in the Feynman integral representation\cite{1961Wells,1963Feynman,RichardPhillipsFeynmanAlbertR.Hibbs2010}.
They have been used since 1959\cite{1961Wells} in this framework,
which later became the basis for some of the first successful methods
for calculating `numerically exact' R$\rho$D in OQSs with large environments.
The most popular model for OQSs is currently a Feynman-Vernon model
(see equation \ref{eq:hamiltonianFV}) with a continuous spectral
distribution $J(\omega)$ (see equation \ref{eq:spectralDistribution}),
in which the entire influence of the OQS's environment on its R$\rho$D
is captured by the influence functional. 

The remainder of this paper will deal with this OQS model, for which
discretized Feynman integrals that use discretized influence functionals
(\textbf{DIF}s) have been used extensively for calculating the R$\rho$D
of various OQSs\cite{1996Makri,Thorwart1997,Kuhn1999}. For this model,
these DIFs are traditionally expressed in terms of improper integrals
of $J(\omega)$ whose integration bounds are infinite in both directions\cite{1995Makri,1995Makri2}.
The accuracy of these integrals can be extremely important when using
the Feynman integral to numerically calculate R$\rho$D in OQSs, and
a converged calculation of them can be very computationally demanding
in some cases. Aside from the inconvenience of having to check for
convergence when numerically calculating these integrals, not having
an analytic expression for the DIF makes it difficult to examine properties
of it, such as its dependence on the parameters of the physical system
and the dependence of its size on the size of the time steps in the
discretization of the Feynman integral.

In this paper we will give a general expression for the DIF, in terms
of proper integrals of the bath response function $\alpha(t)$, instead
of improper integrals of $J(\omega)$ whose integration bounds are
infinite in both directions. As long as $\alpha(t)$ is represented
analytically, these integrals are easy to evaluate analytically, or
if they cannot be, they are still much easier to evaluate numerically
than the integrals in the traditional expression for the DIF, since
their integration bounds are finite and very tiny. 

One particular form for $\alpha(t)$ for which these proper integrals
of $\alpha(t)$ can very easily be evaluated, leading to a fully analytic
DIF%
\footnote{The influence functional is a \textbf{functional} of the Feynman \textbf{paths
}and a regular \textbf{function} of all other independent \textbf{variables}.
When the Feynman paths $s(t)$ are discretized with respect to time,
they can be represented by a set of \textbf{variables }that are constant
with respect to time: $\{s_{k}\}$. A discretized influence functional
can therefore be thought of as a \textbf{functional} of discrete \textbf{paths},
or simply a regular function of a set of \textbf{variables}, and it
is with respect to these variables that the DIF is analytic.%
}, is:

\begin{equation}
\alpha(t)=\sum_{j}^{K}p_{j}e^{\Omega_{j}t}.\label{eq:bathResponseFunctionIsSumOfExponentials}
\end{equation}

This is a very important form for $\alpha(t)$ for at least two reasons. 

(1) Every physically relevant spectral distribution can be fitted
to this form, and in fact many very versatile spectral distribution
forms (including the very popular Lorentz-Drude/Debye function) naturally
have bath response functions of this form (see table \ref{tab:analyticBathResponseFunction}). 

(2) The main purpose of Feynman integrals for R$\rho$D in OQSs is
to benchmark methods that are less computationally expensive. When
benchmarking, it is ideal for both methods to use the exact same\textbf{\uline{
}}form for $\alpha(t)$, and many of the techniques that one may wish
to benchmark actually \textbf{require} $\alpha(t)$ to be in this
form! Some examples include techniques based on the evergreen Nakajima-Zwanzig
equation from late the 1950s\cite{1958Nakajima,1960Zwanzig,Meier1999},
the hierarchical equations of motion (HEOM) which provide argubly
the most successful and robust method for calculating R$\rho$D in
OQSs to date\cite{1989Tanimura,1990Tanimura}, and the very recent
NMQSD-ZOFE quantum master equation\cite{Ritschel2011a}.

For these reasons, we will present analytic expressions for the DIF
for this form of $\alpha(t)$ as an example in section II\ref{sec:Results}.
Table \ref{tab:analyticBathResponseFunction} represents $\alpha(t)$
in the form of equation \ref{eq:bathResponseFunctionIsSumOfExponentials},
with the $p_{j}$ and $\Omega_{j}$ parameters expressed analytically
in terms of the inverse temperature $\beta$ and the parameters of
$J(\omega)$, for three widely used and versatile classes of spectral
distributions, each of which can essentially represent any physically
relevant spectral distribution. The DIF for these spectral distribution
functions can then be obtained in analytic form by substituting these
$p_{j}$ and $\Omega_{j}$ parameters into our analytc expressions
for the DIF for this form of $\alpha(t)$.

Additionally, one of the most widely used forms for $J(\omega)$ is
$J(\omega)\propto\omega^{s}e^{-(\omega/\omega_{c})^{q}}$\cite{Leggett1987}.
Table \ref{tab:analyticBathResponseFunction} also presents an analytic
expression for $\alpha(t)$ for this form of spectral distribution,
for certain values of $s$ when $q=1$. The DIF is then obtained in
analytic form and presented in the supplementary material. Likewise,
the DIF can be obtained in analytic form for various other spectral
distribution forms in the same way when an analytic form for $\alpha(t)$
exists, whether naturally or by a numerical fit.

The open source Mathematica program that supplements this paper takes
in the size of the time step $\Delta t$ in the discretization of
the Feynman paths, and any form of $\alpha(t)$ as input; and outputs
an analytic form for the DIF if possible, and quickly calculates the
DIF numerically otherwise. This program, and the open source MATLAB
program that also supplements this paper, can both readily calculate
the bath response functions and reorganization energies of all four
of the general forms of $J(\omega)$ presented in table \ref{tab:analyticBathResponseFunction}.
The MATLAB program can also take in any $J(\omega),$ and provide
$\alpha(t)$ in the form of equation \ref{eq:bathResponseFunctionIsSumOfExponentials}
by outputting the fitted values of $\{p_{j},\Omega_{j}\}$.

\section{Results\label{sec:Results}}

\subsection{Setting}

The most popular OQS model is currently the Feynman-Vernon model.
In the Feynman-Vernon model, the OQS $s$ is coupled linearly to a
set of quantum harmonic oscillators $Q_{k}$:
\begin{alignat}{1}
H & =H_{{\rm OQS}}+H_{{\rm OQS-bath}}+H_{bath}\\
 & =H_{{\rm OQS}}+\sum_{\kappa}c_{\kappa}sQ+\sum_{\kappa}\big(\textrm{\textonehalf}m_{\kappa}\dot{Q_{\kappa}}^{2}+\text{\textonehalf}m_{\kappa}\omega_{\kappa}^{2}Q_{\kappa}^{2}\big)\,.\label{eq:hamiltonianFV}
\end{alignat}
In most models the quantum harmonic oscillators (\textbf{QHO}s) span
a continuous spectrum of frequencies $\omega_{\kappa}$ and the strength
of the coupling between the QHO of frequency $\omega$ and the OQS
is given by the spectral distribution function $J(\omega)$:

\begin{equation}
J(\omega)=\frac{\pi}{2}\sum_{\kappa}\frac{c_{\kappa}^{2}}{m_{\kappa}\omega_{\kappa}}\delta(\omega-\omega_{\kappa})\,.\label{eq:spectralDistribution}
\end{equation}
For the hamiltonian of the Feynman-Vernon model, the bath response
function $\alpha(t)$ is the following integral transform of $J(\omega)$:

\begin{align}
\alpha(t) & =\frac{1}{\pi}\int_{0}^{\infty}J(\omega)\Big(\coth\Big(\frac{\beta\omega\hbar}{2}\Big)\cos(\omega t)-{\rm i}\sin(\omega t)\Big){\rm d}\omega\label{eq:bathResponseFunction}\\
 & =\frac{1}{2\pi}\int_{-\infty}^{\infty}\frac{J(\omega)\exp\Big(\frac{\beta\omega\hbar}{2}\Big)}{\sinh\Big(\frac{\beta\omega\hbar}{2}\Big)}e^{-{\rm i}\omega t}{\rm {\rm d}}\omega,\, J(-\omega)\equiv J(\omega)\label{eq:bathResponseFunctionAsMakri'sFourierTransform}\\
 & =\frac{1}{\pi}\int_{-\infty}^{\infty}\frac{J(\omega)}{1-\exp(-\beta\omega\hbar)}e^{-{\rm i}\omega t}{\rm d}\omega\,,\, J(-\omega)\equiv J(\omega),\label{eq:bathResponseFunctionAsFourierTransformOfBoseFunction}
\end{align}
where, equation \ref{eq:bathResponseFunctionAsFourierTransformOfBoseFunction}
can be written in terms of the Bose-Einstein distribution function
with $x=-\beta\omega\hbar$:

\begin{equation}
f^{{\rm Bose-Einstein}}(x)=\frac{1}{1-\exp(x)}.\label{eq:boseEinsteinFunction}
\end{equation}

In the Feynman integral formalism for expressing the R$\rho$D given
this hamiltonian, all the information about the influence of $H_{{\rm OQS-bath}}+H_{{\rm bath}}$
on the R$\rho$D of the OQS is completely captured by the (gaussian)
influence phase functional\cite{1963Feynman,RichardPhillipsFeynmanAlbertR.Hibbs2010}:

\begin{align}
\Phi[s(t),s^{\prime}(t)] & =-\int_{0}^{t}\hspace{-1.85mm}\int_{0}^{t^{\prime}}\left(s(t^{\prime})-s^{\prime}(t^{\prime})\right)\left(\alpha(t^{\prime}-t^{\prime\prime})s(t^{\prime\prime})-\alpha^{*}(t^{\prime}-t^{\prime\prime})s^{\prime}(t^{\prime\prime})\right){\rm d}t^{\prime\prime}{\rm d}t^{\prime}\label{eq:influencePhaseExplicit}\\
 & \equiv-\int_{0}^{t}\hspace{-1.85mm}\int_{0}^{t^{\prime}}\bar{\Phi}[s(t^{\prime}),s^{\prime}(t^{\prime}),s(t^{\prime\prime}),s^{\prime}(t^{\prime\prime})]{\rm d}t^{\prime}\,,
\end{align}
which defines the (gaussian) influence functional:

\begin{equation}
F[s(t),s^{\prime}(t)]=\exp\left(\Phi\right)\,.
\end{equation}
In order to numerically calculate a Feynman integral for R$\rho$D,
we discretize the Feynman paths: $\{s(t),s^{\prime}(t)\}\rightarrow\{s_{k}^{+},s_{k}^{-}\}_{k=0}^{N}$,
such that each $s_{k}^{\pm}$ is constant with respect to time.

\subsection{Trotter Splitting}

The simplest way to discretize the Feynman paths is to split them
into intervals of equal duration, known as a Trotter splitting. This
allows us to decompose $\Phi[s(t),s^{\prime}(t)]$ into constituent
double integrals:

\begin{alignat}{1}
\Phi[s(t),s^{\prime}(t)] & =-\sum_{k=0}^{N}\sum_{k^{\prime}=0}^{k-1}\int_{k\Delta t}^{(k+1)\Delta t}\bigg(\int_{k^{\prime}\Delta t}^{(k^{\prime}+1)\Delta t}\bar{\Phi}[s_{k}^{+},s_{k}^{-},s_{k^{\prime}}^{+},s_{k^{\prime}}^{-}]{\rm {\rm d}t^{\prime}}+\int_{k\Delta t}^{t^{\prime}}\bar{\Phi}[s_{k}^{+},s_{k}^{-},s_{k}^{+},s_{k}^{-}]\bigg)dt^{\prime\prime}\label{eq:influenceFunctionalDiscretized}\\
 & =-\sum_{k=0}^{N}\sum_{k^{\prime}=0}^{k-1}(s_{k}^{+}-s_{k}^{-})(\eta_{kk^{\prime}}s_{k^{\prime}}^{+}-\eta_{kk}^{*}s_{k^{\prime}}^{-})\,,
\end{alignat}
where, 

\begin{alignat}{1}
\eta_{kk^{\prime}} & =\int_{k\Delta t}^{(k+1)\Delta t}\hspace{-1.85mm}\int_{k^{\prime}\Delta t}^{(k^{\prime}+1)\Delta t}\alpha(t^{\prime}-t^{\prime\prime}){\rm d}t^{\prime\prime}{\rm d}t^{\prime}\,.\label{eq:etaKK'generalTrotter}\\
\eta_{kk} & =\int_{k\Delta t}^{(k+1)\Delta t}\hspace{-1.85mm}\int_{k\Delta t}^{t^{\prime}}\alpha(t^{\prime}-t^{\prime\prime}){\rm d}t^{\prime\prime}{\rm d}t^{\prime}\label{eq:etaKKgeneralTrotter}
\end{alignat}

At this stage, most traditionally \cite{1995Makri,1995Makri2,1995Makri3},
equation \ref{eq:bathResponseFunctionAsMakri'sFourierTransform} is
inserted into the integrand of equations \ref{eq:etaKK'generalTrotter}
and \ref{eq:etaKKgeneralTrotter}, yielding (when $J(-\omega)\equiv-J(\omega)$):

\begin{align}
\eta_{kk^{\prime}} & =\frac{2}{\pi}\int_{-\infty}^{\infty}\frac{J(\omega)}{\omega^{2}}\frac{\exp(\nicefrac{\beta\hbar\omega}{2})}{\sinh(\nicefrac{\beta\hbar\omega}{2})}\sin^{2}(\nicefrac{\omega\Delta t}{2})e^{-{\rm i}\omega\Delta t(k-k^{\prime})}{\rm d\omega}\,\,,\,0\le k^{\prime}<k\le N\,\,,\,\,{\rm and}\label{eq:etaKK'makri}\\
\eta_{kk} & =\frac{1}{2\pi}\int_{-\infty}^{\infty}\frac{J(\omega)}{\omega^{2}}\frac{\exp(\nicefrac{\beta\hbar\omega}{2})}{\sinh(\nicefrac{\beta\hbar\omega}{2})}(1-e^{-{\rm i}\omega\Delta t}-{\rm i}\Delta t\omega){\rm d}\omega\,\,,\,0\le k\le N\,.
\end{align}
Or alternatively \cite{2011Vagov}, one can insert equation \ref{eq:bathResponseFunction},
yielding:

\begin{alignat}{1}
\eta_{kk^{\prime}} & =2\int_{0}^{\infty}\frac{J(\omega)}{\omega^{2}}\big(1-\cos(\Delta t\omega)\big)\big(\cos(\Delta t\omega(k-k^{\prime}))\coth\bigg(\frac{\hbar\beta\omega}{2}\bigg)-{\rm i}\sin(\Delta t\omega(k-k^{\prime}))\big){\rm d}\omega\,\,,\,0\le k^{\prime}<k\le N\,\,,\,\,{\rm and}\label{eq:etaKK'vagov}\\
\eta_{kk} & =\int_{0}^{\infty}\frac{J(\omega)}{\omega^{2}}\big(1-\cos(\Delta t\omega)\big)\coth\bigg(\frac{\hbar\beta\omega}{2}\bigg)+{\rm i\big(}\sin(\Delta t\omega)-\Delta t\omega\big){\rm d}\omega\,\,,\,0\le k\le N\,\,.\label{eq:etaKKvagov}
\end{alignat}

However, if $\alpha(t)=\sum_{j}^{K}p_{j}e^{-\Omega_{j}t}$ we can
calculate equation \ref{eq:etaKK'generalTrotter} analytically to
get:

\begin{alignat}{1}
\eta_{kk^{\prime}} & =4\sum_{j=1}^{K}\frac{p_{j}}{\Omega_{j}^{2}}\sinh^{2}(\nicefrac{\Omega_{j}\Delta t}{2})e^{\Omega_{j}(k-k^{\prime})\Delta t}\,\,,\,0\le k^{\prime}<k\le N\label{eq:etaKK'whenAlphaIsSumOfExponentials}\\
\eta_{kk} & =2\sum_{j=1}^{K}\frac{p_{j}}{\Omega_{j}^{2}}\bigg(\sinh(\nicefrac{\Omega_{j}\Delta t}{2})e^{\nicefrac{\Omega_{j}\Delta t}{2}}-\frac{1}{2}\Omega_{j}\Delta t\bigg)\,\,,\,0\le k\le N\label{eq:etaKKwhenAlphaIsSumOfExponentials}
\end{alignat}
Figure 1 compares the versions of $\eta_{kk^{\prime}}$ in equations
$\ref{eq:etaKK'makri},\ref{eq:etaKK'vagov}$ and $\ref{eq:etaKK'whenAlphaIsSumOfExponentials}$
as a function of $\Delta k\equiv k-k^{\prime}$ with $\Delta t=0.001{\rm ps}$
, along with the CPU time in Mathematica, for the spectral distribution
$J(\omega)=0.027{\rm ps^{2}}\omega^{3}e^{(-\nicefrac{\omega}{2.2{\rm ps^{-1}}})^{2}}$,
which comes from a recent experimental study of quantum dots\cite{Ramsay2010}
and was used in at least two recent computational studies\cite{McCutcheon2011a,Dattani2012}.
The bath response function for this spectral distribution was fitted
to the form of equation $\ref{eq:bathResponseFunctionIsSumOfExponentials}$
with $K=4$, and the fitted parameters $\{p_{j},\Omega_{j}\}$ are
given in the supplementary material, along with a figure which demonstrates
that there are no discernable differences between the least-squares
fit and the original bath response function obtained by numerical
integration. Figure 1 shows that our analytic form for $\eta_{kk^{\prime}}$
agrees essentially exactly with the other two forms when they are
numerically integrated, but take much less time to compute%
\footnote{The CPU times shown in the figure were measured on a Toshiba Satellite
L300 laptop with an Intel Core 2 Duo CPU T5800 processor with a 2.00
GHz clock rate.%
}. All other analytic $\eta$ coefficients in this paper have been
compared with essentially exact agreement against previous versions
calculated by numerical integration, and the results of these comparisons
are presented in the supplementary material.

\begin{figure}
\caption{Equations \ref{eq:etaKK'makri}, \ref{eq:etaKK'vagov} and \ref{eq:etaKK'whenAlphaIsSumOfExponentials}
are equivalent, but equation \ref{eq:etaKK'whenAlphaIsSumOfExponentials}
can be calculated much more efficiently than the other two. Since
the values of these curves on the ordinate axis depend on the size
of $\Delta t$ (which is a parameter used for numerical computation
of the Feynman integral and not a phsyical characteristic of the system),
all of these curves are scaled by the real part of the funciton when
$\Delta k=0$ in order to have the ordinate axis equal to 1 at most.
Here $J(\omega)=0.027{\rm ps^{2}}\omega^{3}e^{(-\nicefrac{\omega}{2.2{\rm ps^{-1}}})^{2}}$,
which is often used to study quantum dots\cite{2010Ramsay,McCutcheon2011a,Dattani2012},
and $\beta=\frac{1}{k_{B}T}$ with $T=77{\rm K}$. Equation \ref{eq:etaKK'makri}
was originally presented in \cite{1995Makri} and equation \ref{eq:etaKK'vagov}
was originally presented in \cite{2011Vagov}. }
\includegraphics[width=0.9\textwidth]{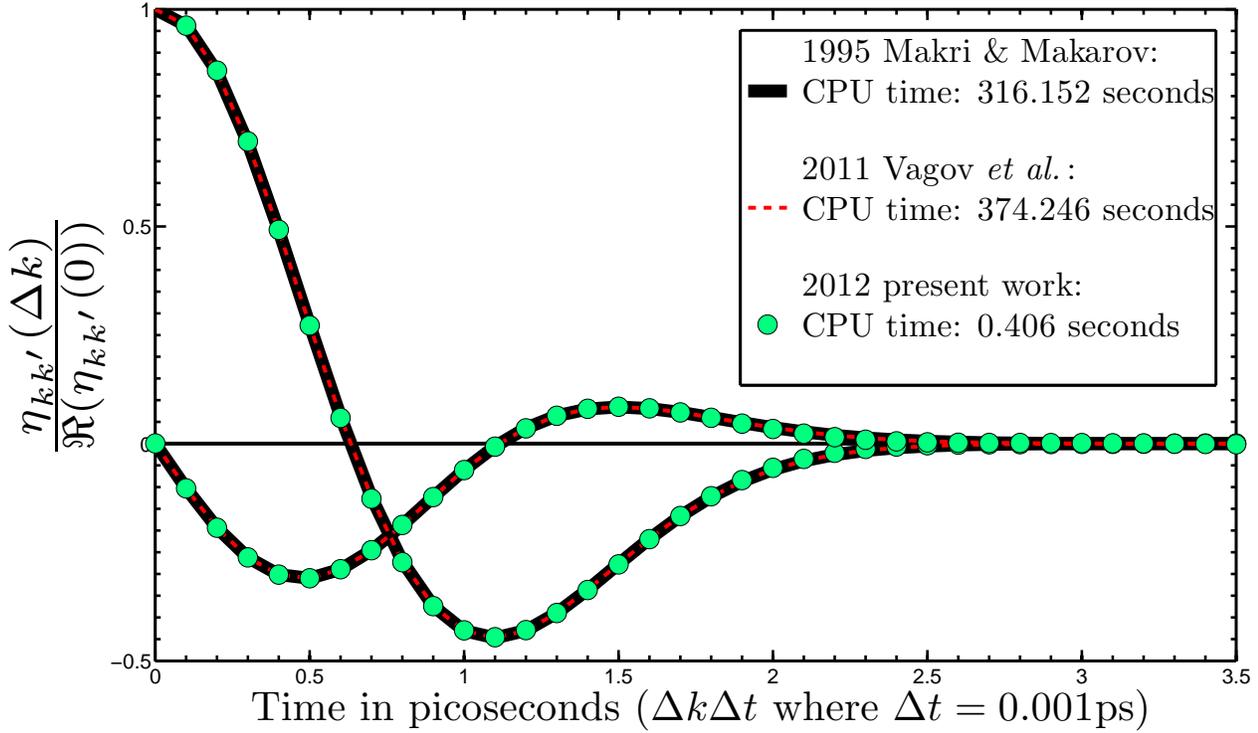}
\end{figure}

\subsection{Strang Splitting}

A more sophisticated discretization of the Feynman paths can be made
by a Strang splitting, which allows one to calculate the R$\rho$D
much more accurately for a given number of time steps in the discretizaton
of the Feynman paths. When this splitting is used, the above formulas
for $\eta_{kk^{\prime}}$ and $\eta_{kk}$ remain the same for $0<k^{\prime}<k<N$,
but different formulas have to be used when either $k$ or $k^{\prime}$,
or both, are $0$ or $N$:

\begin{alignat}{1}
\eta_{N0} & =\int_{N\Delta t-\nicefrac{\Delta t}{2}}^{N\Delta t}\quad\int_{0}^{\nicefrac{\Delta t}{2}}\alpha(t^{\prime}-t^{\prime\prime}){\rm d}t^{\prime\prime}{\rm d}t^{\prime}\label{eq:etaN0generalStrang}\\
\eta_{00} & =\int_{0}^{\nicefrac{\Delta t}{2}}\hspace{-1.85mm}\int_{0}^{t^{\prime}}\alpha(t^{\prime}-t^{\prime\prime}){\rm d}t^{\prime\prime}{\rm d}t^{\prime}\label{eq:eta00generalStrang}\\
\eta_{NN} & =\int_{N\Delta t-\nicefrac{\Delta t}{2}}^{N\Delta t}\quad\int_{N\Delta t-\nicefrac{\Delta t}{2}}^{t^{\prime}}\alpha(t^{\prime}-t^{\prime\prime}){\rm d}t^{\prime\prime}{\rm d}t^{\prime}\label{eq:etaNNgeneralStrang}\\
\eta_{k0} & =\int_{k\Delta t}^{(k+1)\Delta t}\hspace{-1.85mm}\int_{0}^{\nicefrac{\Delta t}{2}}\alpha(t^{\prime}-t^{\prime\prime}){\rm d}t^{\prime\prime}{\rm d}t^{\prime}\label{eq:etaK0generalStrang}\\
\eta_{Nk} & =\int_{N\Delta t-\nicefrac{\Delta t}{2}}^{N\Delta t}\quad\int_{k\Delta t}^{(k+1)\Delta t}\alpha(t^{\prime}-t^{\prime\prime}){\rm d}t^{\prime\prime}{\rm d}t^{\prime}\label{eq:etaNkgeneralStrang}
\end{alignat}

Again, most traditionally \cite{1995Makri,1995Makri2,1995Makri3}
equation \ref{eq:bathResponseFunctionAsMakri'sFourierTransform} is
inserted into equations $\ref{eq:etaN0generalStrang}$ to \ref{eq:etaNkgeneralStrang}
, which yields (when $J(-\omega)\equiv-J(\omega)$):

\begin{alignat}{1}
\eta_{N0} & =\frac{2}{\pi}\int_{-\infty}^{\infty}\frac{J(\omega)}{\omega^{2}}\frac{\exp(\nicefrac{\beta\hbar\omega}{2})}{\sinh(\nicefrac{\beta\hbar\omega}{2})}\sin^{2}(\nicefrac{\omega\Delta t}{4})e^{-{\rm i}\omega(t-\nicefrac{\Delta t}{2})}{\rm d\omega\,\,,}\label{eq:etaN0makri}\\
\eta_{00} & =\eta_{NN}=\frac{1}{2\pi}\int_{-\infty}^{\infty}\frac{J(\omega)}{\omega^{2}}\frac{\exp(\nicefrac{\beta\hbar\omega}{2})}{\sinh(\nicefrac{\beta\hbar\omega}{2})}\left((1-e^{-{\rm i}\omega\nicefrac{\Delta t}{2}})-{\rm \nicefrac{i\Delta t\omega}{2}}\right){\rm d}\omega\,\,,\,0<k<N\label{eq:eta00Makri}\\
\eta_{k0} & =\frac{2}{\pi}\int_{-\infty}^{\infty}\frac{J(\omega)}{\omega^{2}}\frac{\exp(\nicefrac{\beta\hbar\omega}{2})}{\sinh(\nicefrac{\beta\hbar\omega}{2})}\sin(\nicefrac{\omega\Delta t}{4})\sin(\nicefrac{\omega\Delta t}{2})e^{-{\rm i}\omega(k\Delta t-\nicefrac{\Delta t}{4})}{\rm d\omega}\,\,,\,0<k<N,\label{eq:etaK0makri}\\
\eta_{Nk} & =\frac{2}{\pi}\int_{-\infty}^{\infty}\frac{J(\omega)}{\omega^{2}}\frac{\exp(\nicefrac{\beta\hbar\omega}{2})}{\sinh(\nicefrac{\beta\hbar\omega}{2})}\sin(\nicefrac{\omega\Delta t}{4})\sin(\nicefrac{\omega\Delta t}{2})e^{-{\rm i}\omega(t-k\Delta t-\nicefrac{\Delta t}{4})}{\rm d\omega}\,\,,\,0<k<N\label{eq:etaNKmakri}
\end{alignat}

In the case of $\alpha(t)=\sum_{j}^{K}p_{j}e^{\Omega_{j}t}$ we again
find analytical expressions:

\begin{alignat}{1}
\eta_{N0} & =4\sum_{j=1}^{K}\frac{p_{j}}{\Omega_{j}^{2}}e^{\Omega_{j}(t-\nicefrac{\Delta t}{2})}\left(\sinh^{2}(\nicefrac{\Omega_{j}\Delta t}{4})\right)\label{eq:etaN0whenAlphaIsSumOfExponentials}\\
\eta_{00} & =\eta_{NN}=2\sum_{j=1}^{M}\frac{p_{j}}{\Omega_{j}^{2}}\left(e^{\nicefrac{\Omega_{j}\Delta t}{4}}\sinh(\nicefrac{\Omega_{j}\Delta t}{4})-\nicefrac{\Delta t\Omega_{j}}{4}\right)\label{eq:eta00whenAlphaIsSumOfExponentials}\\
\eta_{k0} & =4\sum_{j=1}^{K}\frac{p_{j}}{\Omega_{j}^{2}}\sinh(\nicefrac{\Omega_{j}\Delta t}{2})\sinh\left(\nicefrac{\Omega_{j}\Delta t}{4}\right)e^{\Omega_{j}(k\Delta t-\nicefrac{\Delta t}{4})}\label{eq:etaK0whenAlphaIsSumOfExponentials}\\
\eta_{Nk} & =4\sum_{j=1}^{K}\frac{p_{j}}{\Omega_{j}^{2}}\sinh(\nicefrac{\Omega_{j}\Delta t}{2})\sinh\left(\nicefrac{\Omega_{j}\Delta t}{4}\right)e^{\Omega_{j}(t-k\Delta t-\nicefrac{\Delta t}{4})}\label{eq:etaNKwhenAlphaIsSumOfExponentials}
\end{alignat}

\subsection{Quasi-adiabatic displacement}

When the bath is nearly adiabatic, it is helpful to rewrite equation
\ref{eq:hamiltonianFV} as \cite{1992Makri}:

\begin{alignat}{1}
H & =H_{{\rm OQS}}-H_{{\rm displacement}}+H_{{\rm OQS-bath}}+H_{{\rm bath}}+H_{{\rm displacement}}\\
 & =H_{{\rm OQS}}-\sum_{\kappa}\frac{c_{\kappa}^{2}s^{2}}{2m_{\kappa}\omega_{\kappa}^{2}}+\sum_{\kappa}c_{\kappa}sQ+\sum_{\kappa}\big(\textrm{\textonehalf}m_{\kappa}\dot{Q_{\kappa}}^{2}+\text{\textonehalf}m_{\kappa}\omega_{\kappa}^{2}Q_{\kappa}^{2}\big)+\sum_{\kappa}\frac{c_{\kappa}^{2}s^{2}}{2m_{\kappa}\omega_{\kappa}^{2}}\\
 & \equiv H_{{\rm OQS,displaced}}+\sum_{\kappa}c_{\kappa}sQ+H_{{\rm bath,displaced}}.
\end{alignat}
$H_{{\rm displacement}}$ is called the {}``counter term'', and
can also be represented in terms of the spectral distribution by recognizing
that when the QHOs span a continuous spectrum of frequencies $\omega_{\kappa}$
, we have the relation (remembering equation \ref{eq:spectralDistribution}):

\begin{equation}
\sum_{\kappa}\frac{c_{\kappa}^{2}}{2m_{\kappa}\omega_{\kappa}^{2}}=\frac{1}{\pi}\int_{0}^{\infty}\frac{J(\omega)}{\omega}{\rm d}\omega\,.
\end{equation}

The influence of $H_{{\rm OQS-bath}}+H_{{\rm bath,displaced}}$ on
the R$\rho$D of the displaced OQS is now completely captured by the
modified influence phase functional\cite{1995Makri,1995Makri2,1995Makri3}:

\begin{flushleft}
{\small 
\begin{equation}
\Phi_{{\rm QUAPI}}[s(t),s^{\prime}(t)]=-\int_{0}^{t}\hspace{-1.85mm}\int_{0}^{t^{\prime}}\bigg(\left(s(t^{\prime})-s^{\prime}(t^{\prime})\right)\left(\alpha(t^{\prime},t^{\prime\prime})s(t^{\prime\prime})-\alpha^{*}(t^{\prime},t^{\prime\prime})s^{\prime}(t^{\prime\prime})\right)+\frac{{\rm i}}{\hbar\pi}\int_{0}^{t}\int_{0}^{\infty}\frac{J(\omega)}{\omega}{\rm d}\omega\Big(s^{+}(t^{\prime})^{2}-s^{-}(t^{\prime})^{2}\Big){\rm \bigg)d}t^{\prime\prime}{\rm d}t^{\prime}\,,
\end{equation}
}where QUAPI stands for \textbf{Qu}asi\textbf{-A}diabatic\textbf{
P}ropagator Feynman%
\footnote{The term `path integral' is used more commonly than `Feynamn integral'
here, but this term is ambiguous. Currently, the first result on the
search engine at www.google.com, when the search query `path integral'
is entered, is a Wikipedia page that currently links to three different
meanings of the word `path integral': (1) line integral, (2) functional
integration, and (3) path integral formulation. Only the third of
these is unambiguously the Feynman integral discussed in this paper.
The `line integral' is an integral over a path, rather than over a
set of paths; and the term `functional integral' can refer to at least
three types of functional integrals: (1) the Wiener integral, (2)
the Lévy integral, and (3) the Feynman integral.%
} \textbf{I}ntegral, which is the name for the resulting Feynman integral
for the R$\rho$D of the displaced OQS when this new influence phase
functional is used. When the bath is nearly adiabatic, the QUAPI is
more accurate than a Feynman integral that does not make use of this
modified influence phase functional, for a given size of the time
step used in the discretizaton of the Feynman paths. The new term
in the influence phase functional adds the following term to the $\eta_{kk^{\prime}}$
coefficients when $k=k^{\prime}$: 
\par\end{flushleft}

\begin{flushleft}
\begin{alignat}{1}
\asterism_{{\rm QUAPI}} & =\frac{{\rm i}\Delta t}{\hbar\pi}\int_{0}^{\infty}\frac{J(\omega)}{\omega}{\rm d}\omega\label{eq:QUAPItermInInfluenceFunctionalWRTspectralDistribution}\\
 & \equiv\frac{{\rm i}\Delta t}{\hbar\pi}\lambda\,,
\end{alignat}
where in the last line we have defined the {}``bath reorganization
energy'' by $\lambda$. In equations \ref{eq:etaKK'makri} to \ref{eq:etaKKvagov},
and \ref{eq:etaN0makri} to \ref{eq:etaNKmakri}, equation $\ref{eq:QUAPItermInInfluenceFunctionalWRTspectralDistribution}$
can simply be incorporated into the integrals over $\omega$, but
our versions (equations \ref{eq:etaKK'whenAlphaIsSumOfExponentials},
\ref{eq:etaKKwhenAlphaIsSumOfExponentials}, and \ref{eq:etaN0whenAlphaIsSumOfExponentials}
to \ref{eq:etaNKwhenAlphaIsSumOfExponentials}) are no longer analytic
unless an analytic expression exists for $\lambda$. Fortunately an
analytic expression exists for $\lambda$ for most forms of $J(\omega)$.
Table \ref{tab:analyticBathResponseFunction} presents analytic expressions
for some of the most popular forms of $J(\omega)$.
\par\end{flushleft}

\subsection{New class of spectral distributions whose bath response functions
are analytically expressed as $\alpha(t)=\sum_{j}^{K}p_{j}e^{\Omega_{j}t}$.}

This formula is good for testing the spectral distribution corresponding
to a bath response function obtained by a least-squares fit, or obtained
by a truncation of the (exact) infinite series expression for $\alpha(t)$
table (see table \ref{tab:analyticBathResponseFunction} for some
examples).

\begin{equation}
J(\omega)=\Re\frac{1}{\pi}(1-e^{-\beta\omega})\sum_{j}^{K}\frac{{\rm i}p_{j}}{\omega-{\rm i}\Omega_{j}}\label{eq:manolopoulosTransform}
\end{equation}

\section{Conclusions}

We have presented general formulas for the $\eta$ coefficients in
the discretized influence functional of the Feynman-Vernon model,
in terms of the bath response function. These formulas can then be
used to derive the form of the $\eta$ coefficients presented by Makarov
and Makri in 1995\cite{1995Makri,1995Makri2}, and the form presented
by Vagov $\textit{et al.}$ in 2011\cite{2011Vagov}, both which involve
improper integrals of the spectral distribution and have at least
one infinite integration bound. However, when the bath response function
can be expressed analytically (which is almost always the case, as
shown in table \ref{tab:analyticBathResponseFunction}), our general
formulas for the $\eta$ coefficients can almost always be derived
analytically too. Figure 1 shows that the evaluation of these analytic
formulas can be more than 900 times faster than obtaining the exact
same results by numerically integrating forms for the $\eta$ coefficients
in the traditional ways.

If there is a case where the bath response function can be expressed
analytically, but the integrals in equation \ref{eq:etaKK'generalTrotter},
\ref{eq:etaKKgeneralTrotter} and/or \ref{eq:etaN0generalStrang}
to \ref{eq:etaNkgeneralStrang} cannot be evaluated analytically (we
do not know of such a case, but cannot rule out the possibility that
it exists), we recommend to numerically integrate $\textit{these}$
equations, rather than to numerically integrate the more traditional
forms presented in equations \ref{eq:etaKK'makri} to \ref{eq:etaKKvagov}
and \ref{eq:etaN0makri} to \ref{eq:etaNKmakri}. This is because
these integrals have finite bounds, which are very small (about the
size of the time steps in the discretization of the Feynman paths),
while the more traditional forms involve numerically integrating the
spectral distribution over infinite bounds.

When the bath response function $\textit{cannot}$ be expressed analytically,
it may still be more convenient to obtain it by numerically integrating
equation \ref{eq:bathResponseFunction}, \ref{eq:bathResponseFunctionAsMakri'sFourierTransform}
or \ref{eq:bathResponseFunctionAsFourierTransformOfBoseFunction},
and then to fit it to a sum of complex-weighted complex exponentials.
This is because Feynman integrals for calculating the reduced density
operator dynamics in open quantum systems are usually used for benchmarking
other techniques, and many popular techniques \textbf{require }this
form for the bath response function, and when benchmarking it is always
best to keep as many parameters of the model consistent across all
methods being benchmarked. When fitting a bath response function to
this form, equation \ref{eq:manolopoulosTransform} is useful for
testing the quality of the fit in terms of how well the fitted bath
response function corresponds to the original desired spectral distribution.
This formula is equally useful in cases when the bath response function
is naturally represented in the form of equation \ref{eq:bathResponseFunctionIsSumOfExponentials}
(as in the examples in table \ref{tab:analyticBathResponseFunction}),
since it can give an indication of how many terms in the infinite
series are required in order to be sure that one is using a bath response
function which corresponds to a spectral distribution similar enough
to the original desired one.

\section{Special Cases (Appendix)}

The results in equations \ref{eq:etaKK'whenAlphaIsSumOfExponentials},
\ref{eq:etaKKwhenAlphaIsSumOfExponentials} and \ref{eq:etaN0whenAlphaIsSumOfExponentials}
to \ref{eq:etaNKwhenAlphaIsSumOfExponentials} can readily be applied
for many of the spectral distributions listed in table \ref{tab:analyticBathResponseFunction},
by substituting the appropriate values of $p_{j}$ and $\Omega_{j}$.

One of the most popular spectral distribution forms is currently the
Lorentz-Drude/Debye (LDD) form which has been used extensively in
experimental determinations of spectral distributions \cite{2006Zigmantas,2007Read,2008Read},
and in various OQS studies (\cite{2009Ishizaki} and various other
studies of this same system that compare to this result, including
ones involving DIFs for Feynman integrals \cite{2011Nalbach,Nalbach2011}):

\begin{equation}
J(\omega)=\frac{\omega}{\pi}\frac{\lambda\gamma}{\gamma^{2}+\omega^{2}}.\label{eq:spectralDensityLDD}
\end{equation}
The first spectral distribution form presented in table \ref{tab:analyticBathResponseFunction}
is a generalization of this LDD form, which allows for multiple peaks
to be included at various locations defined by $\{\tilde{\omega}_{j}\}$.
In the case where only one term in the sum exists, and $\tilde{\omega}=0$,
equation $\ref{eq:spectralDensityLDD}$ is recovered. The bath response
function for equation $\ref{eq:spectralDensityLDD}$ can be obtained
analytically in the form of equation \ref{eq:bathResponseFunctionIsSumOfExponentials}
by a simple application of residue calculus, leading to the well-known
Matsubara series. Since the Matsubara series converges extremely slowly,
we have presented a series based on the very recent {[}N-1/N{]} Padé
decomposition technique in \cite{Hu2010} (a more clever application
of residue calculus), which is also of the form in equation \ref{eq:bathResponseFunctionIsSumOfExponentials},
but converges with much fewer terms. The label {[}N-1/N{]} denotes
that the Padé approximant is a rational funciton where the numerator
is a polynomial of degree N-1, and the denominator is a polynomial
of degree N. N effectively becomes the number of exponential terms
in $\alpha(t)$ that specifically arise due to this technique for
expressing $\alpha(t)$ in the form of equation \ref{eq:bathResponseFunctionIsSumOfExponentials}
(more details about this technique can be found in \cite{Hu2010,Hu2011}).%
\footnote{N is therefore different from $N$ that was used throughout this paper
to denote the number of time steps used in the discretization of the
Feynman paths.%
} \textbf{This representation of $\alpha(t)$ becomes more and more
accurate as $N$ is increased.} The Padé parameters $\{\xi_{n}$,$\Xi_{n}\}$
are given in table \ref{tab:[N-1/N]-Pad=0000E9-parameters}. The total
number of terms in $\alpha(t)$ is given by $K=2J+N$, where $J$
is the number lorentzian terms of each type in $J(\omega)$.

The second spectral distribution form presented in the table below
is used in studies where the $J(\omega)$ for a system is predicted
using a molecular dynamics simulation\cite{Damjanovic2002,Olbrich2011a}.
It is the same as the first spectral distribution form in table \ref{tab:analyticBathResponseFunction},
but with the prefactor $\omega$ replaced by $\tanh(\frac{\beta\omega}{2})$
in an attempt to remove the temperature dependence of $J(\omega)$
which is inherent due to the nature of the molecular dynamics technique
that is used to calculate it. As for the first $J(\omega)$ presented,
the bath response function for this $J(\omega)$ was put into the
form of equation \ref{eq:bathResponseFunctionIsSumOfExponentials}
by the Padé decomposition, except that in this case only the imaginary
component requires a Padé decomposition treatment. The total number
of terms in $\alpha(t)$ is again given by $K=2J+N$.

The third spectral distribution form presented is known as the Meier-Tanor
decomposition and has been widely used since its introduction in 1999\cite{Meier1999}.
It is similar to forms used for the spectral distribution function
in \cite{Shugard1978} and \cite{Thorwart2004}. Here we do not use
a Padé decomposition treatment, but rather present the same representation
of $\alpha(t)$ in the form of equation \ref{eq:bathResponseFunctionIsSumOfExponentials}
as was originally presented by Meier and Tanor (using a Matsubara-style
decomposition) in \cite{Meier1999}. The total number of terms in
$\alpha(t)$ is again given by $K=2J+N$, where $N$ now denotes the
number of terms contributed by the Matsubara-style decomposition.
Once again, this representation of $\alpha(t)$ beocmes more and more
accurate as $N$ is increased.

Finally, formulas \ref{eq:etaKK'generalTrotter}, \ref{eq:etaKKgeneralTrotter}
and \ref{eq:etaN0generalStrang} to \ref{eq:etaNkgeneralStrang} can
readily be applied to other spectral distribution functions when an
analytic expression for $\alpha(t)$ exists. One very popular class
of spectral distribution functions for which $\alpha(t)$ can be calculated
analytically (resulting in a form different from equation \ref{eq:bathResponseFunctionIsSumOfExponentials})
is the class of functions of the form $J(\omega)\propto\omega^{s}e^{-(\nicefrac{\omega}{\omega_{c}})^{q}}.$
An analytic form for $\alpha(t)$ and the reorganization energy $\lambda$
for this class of spectral distribution functions is presented in
table \ref{tab:analyticBathResponseFunction} below. The result of
evaluating this formula is compared to the $\alpha(t)$ obtained by
numerical integration with essentially exact agreement, and this comparison
is presented in the supplementary material.

\begin{table}
\caption{Analytic formulas for the bath response functions $\alpha(t)$ and
reorganization energies $\lambda$ of various selected spectral distribution
functions. For the first three cases, $K=2J+N$, where $N$ is the
number of terms in $\alpha(t)$ arising from the Padé (for the first
two forms of $J(\omega)$) or Matsubara-style (for the third form
of $J(\omega)$) decomposition. All formulas presented here can be
computed easily for any value of $K$ in our open source MATLAB program
that supplements this paper. %
\begin{lyxgreyedout}
$\psi_{s}(\cdot)$ is the polygamma function, which is a built-in
function that is evaluated quickly in MATLAB, Mathematica, and most
computing software.%
\end{lyxgreyedout}
{} \label{tab:analyticBathResponseFunction}}

{\footnotesize }%
\begin{tabular}{lcr}
\hline 
{\footnotesize $J(\omega)=\frac{\omega}{\pi}\sum_{j}^{J}\bigg(\frac{\lambda{}_{j}\gamma_{j}}{\gamma_{j}^{2}+(\omega-\tilde{\omega}_{j})^{2}}+\frac{\lambda{}_{j}\gamma_{j}}{\gamma_{j}^{2}+(\omega+\tilde{\omega}_{j})^{2}}\bigg)$ } & {\footnotesize $\alpha(t)=\sum_{j}^{K}p_{j}e^{\Omega_{j}t}$} & \tabularnewline[5mm]
\hline 
\noalign{\vskip2mm}
\noalign{\vskip2mm}
\multicolumn{2}{l}{{\footnotesize $\quad\quad p_{j=}\begin{cases}
\frac{\lambda{}_{j}}{\beta}\left(1-\sum_{n=1}^{N}\frac{2\Xi_{n}\Omega_{j}^{2}}{\nu_{n}^{2}-\Omega_{j}^{2}}\right)+{\rm i}\frac{\lambda{}_{j}\Omega_{j}}{2}\,, & j\in[0,J]\\
\frac{\lambda{}_{j}}{\beta}\left(1-\sum_{n=1}^{N}\frac{2\Xi_{n}\Omega_{j}^{2}}{\nu_{n}^{2}-\Omega_{j}^{2}}\right)+{\rm i}\frac{\lambda{}_{j}\Omega_{j}}{2}\,, & j\in[J+1,2J]\\
-\sum_{m=1}^{J}\frac{4\lambda_{m}\gamma_{m}}{\beta}\frac{\Xi_{j}\nu_{j}\big(|\Omega_{m}|^{2}-\nu_{j}^{2}\big)}{\lvert\nu_{j}^{2}-\Omega_{m}^{2}\rvert^{2}}\,, & j\in[2J+1,K]
\end{cases}$}} & \tabularnewline
\noalign{\vskip2mm}
\multicolumn{2}{l}{{\footnotesize $\hspace{1em}\hspace{1em}\Omega_{j}=\begin{cases}
-\gamma{}_{j}+{\rm i}\tilde{\omega}_{j}\,, & j\in[0,J]\\
-\gamma{}_{j}-{\rm i}\tilde{\omega}_{j}\,, & j\in[J+1,2J]\\
-\nu_{j}\,, & j\in[2J+1,K]
\end{cases}$}} & {\footnotesize ${\color{white}\begin{cases}
{\color{black}}\\
{\color{black}(\cdot)_{J+j}} & {\color{black}{\color{black}=(\cdot)_{j}}}\\
{\normalcolor \nu_{n}} & {\normalcolor =\frac{2\pi\xi_{n}}{\beta}}
\end{cases}}$}\tabularnewline
\noalign{\vskip2mm}
{\footnotesize $\hspace{1em}\hspace{1em}\lambda=\sum_{j}^{J}\lambda_{j}$} & \multicolumn{2}{r}{\textcolor{black}{\footnotesize $\{\xi_{n},\Xi_{n}\}$ are the {[}N-1/N{]}
Padé parameters for the Bose-Einstein distribution (see table \ref{tab:[N-1/N]-Pad=0000E9-parameters})}}\tabularnewline
\noalign{\vskip2mm}
\hline 
\noalign{\vskip5mm}
{\footnotesize $J(\omega)=\frac{1}{\pi}\tanh\big(\frac{\beta\omega}{2}\big)\sum_{j}^{J}\bigg(\frac{\lambda_{j}\gamma_{j}}{\gamma_{j}^{2}+(\omega-\tilde{\omega}_{j})^{2}}+\frac{\lambda_{j}\gamma_{j}}{\gamma_{j}^{2}+(\omega+\tilde{\omega}_{j})^{2}}\bigg)$} & {\footnotesize $\alpha(t)=\sum_{j}^{K}p_{j}e^{\Omega_{j}t}$} & \textcolor{black}{\footnotesize $J(\omega)$ introduced by Damjanovic
$\textit{et al.}$ \cite{Damjanovic2002,Gutierrez2010}}\tabularnewline[5mm]
\hline 
\noalign{\vskip2mm}
\noalign{\vskip2mm}
\multicolumn{2}{l}{{\footnotesize $\hspace{1em}\hspace{1em}p_{j}=\begin{cases}
\frac{\lambda_{j}}{2}+{\rm i}\frac{2\lambda_{j}}{\beta}\sum_{n=1}^{N}\frac{\Xi_{n}\Omega_{j}}{\nu_{n}^{2}-\Omega_{j}^{2}}\,, & j\in[0,J]\\
\frac{\lambda_{j}}{2}+{\rm i}\frac{2\lambda_{j}}{\beta}\sum_{n=1}^{N}\frac{\Xi_{n}\Omega_{j}}{\nu_{n}^{2}-\Omega_{j}^{2}}\,, & j\in[J+1,2J]\\
-{\rm i}\frac{4\Xi_{j}}{\beta}\sum_{m=1}^{J}\frac{\tilde{\gamma}_{m}\tilde{\lambda}_{m}\big(|\Omega_{m}|^{2}-\nu_{j}^{2}\big)}{\lvert\nu_{j}^{2}-\Omega_{m}^{2}\rvert^{2}}\,, & j\in[2J+1,K]
\end{cases}$ }} & \tabularnewline
\noalign{\vskip2mm}
\multicolumn{2}{l}{{\footnotesize $\hspace{1em}\hspace{1em}\Omega_{j}=\begin{cases}
-\gamma_{j}+{\rm i}\tilde{\omega}_{j}\,, & j\in[0,J]\\
-\gamma_{j}-{\rm i}\tilde{\omega}_{j}\,, & j\in[J+1,2J]\\
-\nu_{j}\,, & j\in[2J+1,K]
\end{cases}$}} & {\footnotesize ${\color{white}\begin{cases}
{\color{black}}\\
{\color{black}(\cdot)_{J+j}} & {\color{black}{\color{black}=(\cdot)_{j}}}\\
{\normalcolor \nu_{n}} & {\normalcolor =\frac{2\pi\xi_{n}}{\beta}}
\end{cases}}$}\tabularnewline
\noalign{\vskip2mm}
 & \multicolumn{2}{r}{\textcolor{black}{\footnotesize $\{\xi_{n},\Xi_{n}\}=$are the {[}N-1/N{]}
Padé parameters for the Fermi-Dirac distribution (see table \ref{tab:[N-1/N]-Pad=0000E9-parameters})}}\tabularnewline
\noalign{\vskip2mm}
\hline 
\noalign{\vskip5mm}
{\footnotesize $J(\omega)=\frac{\pi\omega}{2}\sum_{j}^{K}\frac{\lambda_{j}}{\big(\gamma_{j}^{2}+(\omega+\tilde{\omega}_{j})^{2}\big)\big(\gamma_{j}^{2}+(\omega-\tilde{\omega}_{j})^{2}\big)}$} & {\footnotesize $\alpha(t)=\sum_{j}^{K}p_{j}e^{\Omega_{j}t}$} & {\footnotesize Meier-Tanor decomposition \cite{Meier1999}}\tabularnewline[5mm]
\hline 
\noalign{\vskip2mm}
\noalign{\vskip2mm}
\multicolumn{2}{l}{{\footnotesize $\quad\quad p_{j=}\begin{cases}
\frac{\lambda j}{\tilde{\omega}_{j}\gamma_{j}}\bigg(\coth\big(\frac{\beta}{2}(\tilde{\omega}_{j}+{\rm i}\gamma_{j})\big)+{\rm i}\bigg)\,, & j\in[0,J]\\
\frac{\lambda_{j}}{\tilde{\omega}_{j}\gamma_{j}}\bigg(\coth\big(\frac{\beta}{2}(\tilde{\omega}_{j}-{\rm i}\gamma_{j})\big)-{\rm i}\bigg)\,, & j\in[J+1,2J]\\
-\frac{\pi}{\beta}\sum_{n=1}^{N}\frac{\nu_{j}\lambda_{n}}{\big(\gamma_{n}^{2}+(i\nu_{j}+\tilde{\omega}_{n})^{2}\big)\big(\gamma_{n}^{2}+(i\nu_{j}-\tilde{\omega}_{n})^{2}\big)}\,, & j\in[2J+1,K]
\end{cases}$}} & \tabularnewline
\noalign{\vskip2mm}
\multicolumn{2}{l}{{\footnotesize $\hspace{1em}\hspace{1em}\Omega_{j}=\begin{cases}
-\gamma_{j}+{\rm i}\tilde{\omega}_{j}\,, & j\in[0,J]\\
-\gamma_{j}-{\rm i}\tilde{\omega}_{j}\,, & j\in[J+1,2J]\\
-\nu_{j}\,, & j\in[2J+1,K]
\end{cases}$}} & {\footnotesize ${\color{white}\begin{cases}
{\color{black}}\\
{\color{black}(\cdot)_{J+j}} & {\color{black}{\color{black}=(\cdot)_{j}}}\\
{\normalcolor \nu_{n}} & {\normalcolor =\frac{2\pi n}{\beta}}
\end{cases}}$}\tabularnewline
\noalign{\vskip2mm}
\multicolumn{3}{l}{{\footnotesize $\hspace{1em}\hspace{1em}\lambda=\frac{\pi^{2}}{8\gamma(\gamma^{2}+\omega_{0}^{2})}\sum_{j}^{K}\lambda_{j}$}}\tabularnewline
\noalign{\vskip2mm}
\hline 
\noalign{\vskip5mm}
{\footnotesize $J(\omega)=A\omega^{s}e^{-\nicefrac{\omega}{\omega_{c}}}$} & {\footnotesize $\alpha(t)=-s!\omega_{c}^{s+1}\left(z+(-1)^{s+1}z^{*}\right)$} & \textcolor{black}{\footnotesize See \cite{Leggett1987} for a detailed
use of this $J(\omega)$}\tabularnewline
\noalign{\vskip2mm}
\multicolumn{3}{c}{{\footnotesize $z\equiv\big(\frac{1}{(\omega_{c}t+i)^{s+1}}\big)-\frac{2\Re\big(\psi_{s}(\frac{1+i\omega_{c}t}{\beta\omega_{c}})\big)}{s!(\beta\omega_{c})^{s+1}}-i\frac{\sin\big((s+1)\arctan(\omega_{c}t)\big)}{(1+\omega_{c}^{2}t^{2})^{(s+1)/2}}$}}\tabularnewline[5mm]
\hline 
\noalign{\vskip2mm}
\hline 
\noalign{\vskip2mm}
\multicolumn{3}{l}{{\footnotesize $\hspace{1em}\hspace{1em}\lambda=A\omega_{c}^{s}\Gamma(s)$}}\tabularnewline
\noalign{\vskip2mm}
\hline 
\multicolumn{3}{l}{{\footnotesize $J(\omega)=A\omega^{s}e^{-(\nicefrac{\omega}{\omega_{c}})^{q}}\, s>0\,,\,\omega_{c}>0\,,\, q>0$}}\tabularnewline
\hline 
\noalign{\vskip2mm}
\multicolumn{3}{l}{{\footnotesize $\hspace{1em}\hspace{1em}\lambda=\frac{A}{q}\omega_{c}^{s}\Gamma(\frac{s}{q})$}}\tabularnewline
\noalign{\vskip2mm}
\end{tabular}
\end{table}

\begin{table}
\caption{{[}N-1/N{]} Padé parameters for the Bose-Einstein and Fermi-Dirac
distribution functions (first presented in \cite{Hu2010} and in more
detail in \cite{Hu2011}). $\{\Xi_{n},\xi_{n}\}$ can be calculated
easily for all $n$, for arbitrary values of $N$ in our open source
MATLAB program that supplements this paper. The matrix $\Lambda$
is a 2N$\times$2N, and $\tilde{\Lambda}$ is a 2N-1$\times$2N-1.
The indices $n$ run from 1 to half the number of non-zero eigenvalues
of the corresponding matrix. %
\begin{lyxgreyedout}
The number of non-zero eigenvalues will always be even, because the
eigenvalues of these particular matrices come in pairs, for example
($\xi_{n},-\xi_{n}$); and for matrices with an odd number of eigenvalues,
the unpaired eigenvalue will always be 0.%
\end{lyxgreyedout}
\label{tab:[N-1/N]-Pad=0000E9-parameters}}

\begin{tabular}{c|c}
\multicolumn{1}{c}{$\{\pm\xi_{n}\}=\text{eigenvalues}(\Lambda^{-1})$} & $\{\pm\zeta_{n}\}=\text{eigenvalues}(\tilde{\Lambda}^{-1})$\tabularnewline
\hline 
\hline 
Bose-Einstein & Fermi-Dirac\tabularnewline
$\Lambda_{cd}=\frac{\delta_{c,d\pm1}}{2\sqrt{(2c+1)(2d+1)}}$ & $\Lambda_{cd}=\frac{\delta_{c,d\pm1}}{2\sqrt{(2c-1)(2d-1)}}$\tabularnewline
$\tilde{\Lambda}_{cd}=\frac{\delta_{c,d\pm1}}{2\sqrt{(2c+3)(2d+3)}}$ & $\tilde{\Lambda}_{cd}=\frac{\delta_{c,d\pm1}}{2\sqrt{(2c+1)(2d+1)}}$\tabularnewline
$\Xi_{n}=(N^{2}+\frac{3}{2}N)\frac{\prod_{m=1}^{N-1}(\zeta_{m}^{2}-\xi_{n}^{2})}{\prod_{m\neq n}^{N}(\xi_{m}^{2}-\xi_{n}^{2})}$ & $\Xi_{n}=(N^{2}+\frac{1}{2}N)\frac{\prod_{m=1}^{N-1}(\zeta_{m}^{2}-\xi_{n}^{2})}{\prod_{m\neq n}^{N}(\xi_{m}^{2}-\xi_{n}^{2})}$\tabularnewline
\end{tabular}
\end{table}

\newpage{}

\section{Acknowledgements}

We would like to thank Thomas E. Markland whose MSc. thesis inspired
this work. For valuable comments on the manuscript, and for the derivation
of equation \ref{eq:manolopoulosTransform}, credit is due to David
E. Manolopoulos. N.S.D. thanks the Clarendon Fund and the NSERC/CRSNG
of/du Canada for financial support. F.A.P. thanks the Leverhulme Trust
for financial support. We dedicate this manuscript to Albert Einstein
and Brendon W. Lovett, who were both born on this day.

\bibliographystyle{plain}

\end{document}